# Hydrophobic Silica Microcavities with Sustainable Nonlinear Photonic Performance

Jiadu Xie, Yang Wang, Hui Kang, Jinsong Cheng, Xiaoqin Shen*

School of Physical Science and Technology, ShanghaiTech University, Shanghai, 201210, China

Email: shenxq@shanghaitech.edu.cn

**Abstract**: Ultrahigh quality factor (Q) microcavities have been emerging as an appealing compact photonic platform for various applications. The Q factor plays a critical role in determining the nonlinear optical performance of a microcavity. However, a silica microcavity suffers from severe degradation of its Q value over time during storage or use in air due to the accumulating surface absorption loss, which would deteriorate their nonlinear photonic performance. Here, we report a new type of ultrahigh Q silica microcavity that effectively prevents the Q degradation over time. The Q values of the devices remain unchanged over time under storage in air. Optical frequency combs are generated with sustainable ultralow threshold performance in the course of time from the devices in open air. This approach would greatly facilitate ultrahigh Q silica-based photonic devices for next generation photonic applications.

## Introduction

Whispering gallery mode optical microcavities with high quality factors (Q) have been emerging as appealing on-chip platforms for revolutionizing many photonic technologies due to their ability to trap light efficiently. [1-3] The high Q value promotes the build-up of exceptionally high circulating intensity inside a cavity, enabling various nonlinear optical processes in a compact device. The Q factor plays a critical role in determining the nonlinear optical performance of a microcavity, such as the threshold power,[4] lasing linewidth,[5] and frequency noise.[6] Over the past two decades, efforts have been made to explore high Q microcavities on a range of material platforms,[7] including fused silica,[8-10] silicon nitride,[11] lithium niobates,[12, 13] aluminum nitride[14] and III-V compounds.[15] However, except for silica, most of these material platforms require sophisticated fabricating recipes and expensive equipment to fabricate high Q microcavities. Moreover, due to their intrinsic material properties, the Q values are limited in the range of $10^5$ to $10^7$. Such a moderate high Q device requires a high pump power over several hundred miliwatt to perform,[16] which is still a big challenge for on-chip light source applications.

For silica, owing to the ultra-low intrinsic optical loss, it offers the highest Q values for chip-based system.[9] Recently, either off-chip and on-chip silica microcavity devices with optimized Q values above $1\times10^9$ were achieved by simple laser machining[17] or standard wet etching processes.[18] The ultrahigh Q values and simple fabrication processes make them promising for various nonlinear photonic applications.[3, 6, 10, 19-22] However, the use of air as a cladding layer in silica microcavities makes them susceptible to water molecule absorption on the silica surface, which would bring additional optical loss.[23] It can cause severe degradation of the Q value of a freshly prepared ultrahigh Q silica device over time by more than one order of magnitude.[17] This not only leads to a quadratically increased pump threshold for various nonlinear processes but also affects the cavity opto-thermal instability and overturns the nonlinear photonic performance.[24] Thus, the Q stability issue remains a significant challenge in the applications of silica-based microcomb technologies.

Here, we report a new type of ultrahigh Q silica microcavities that prevents the Q degradation over time, and enables their sustainable nonlinear photonic performances in air for long term. The devices are prepared by grafting a dense monolayer of hydrophobic *1H,1H,2H,2H*-perfluorooctyl (PFO) molecules on the silica microcavity surfaces. The hydrophobic PFO molecular layer effectively inhibits both high affinity and low affinity surface water absorption. Notably, the devices show exceptional long-term Q stability in open air with 35% or even 75% humilities at room temperature. Optical frequency combs with ultralow power thresholds of about 117 μW are generated from the devices in open air by a free running continuous wave laser at 1550 nm. The performances of optical frequency comb in the course of time are highly stable for devices stored over nine months.

**Results and Discussion**

Typical silica microcavities include spherical, toroidal, edge and rod microcavities (**Figure 1a**) can achieve ultrahigh Q values over $1\times10^8$. During storage under ambient condition, they all suffer from Q degradation to different extent over times (**Figure S1, supporting information**). The Q degradation was ascribed to the accumulating surface optical loss caused by the absorption of water molecules on the surface. Water absorptions on a hydrophilic silica surface take place either through high affinity hydrogen bonding interaction between surface hydroxyl groups and water molecules, or through low affinity hydrogen bonding interaction between surface exposed oxygen atoms and water molecules[25, 26] (**Figure 1b, i**). To fully prevent the Q degradation over times caused by water absorption, both high affinity and low affinity hydrogen-bonding interaction should be prevented.

Hydrophobic silica micorcavities are prepared by grafting a monolayer of hydrophobic *1H,1H,2H,2H*-perfluorooctyl (PFO) molecules on to the surface of silica devices, via a self-assembled monolayer approach.[27] In this approach, the chemical vapor of the PFO precursors (*1H,1H,2H,2H*-perfluorooctyl trichlorosilane) first absorb on the surface and in situ react with the surface hydroxyl groups to form a self-assembled monolayer of PFO molecules that covalently bond to the surface (**Figure S2, supporting information**). X-ray photoelectron spectra of the devices clearly show the primary binding energy peak and its Auger peak of F (1s) at 687 eV and 832 eV, respectively, confirming the formation of PFO molecular layer on surface (**Figure S3, supporting information**). Atomic force microscopy (AFM) results confirm the formation of a uniform monolayer on the silica surface. The measured thickness of the layer is of about 0.99 nm (**Figure S4, supporting information**), which is consistent with the thickness of one monolayer of PFO molecules (the theoretical molecular length of one straight PFO molecule is about 1.01 nm).

To evaluate the hydrophilicities of the devices, the surface contact angles of silica devices are monitored. Silicon slides with thermal oxide layer are used for the measurement. For initial silica, the surface contact angle is about 22°, indicating its high hydrophilic nature owing to the existence of surface hydroxyl groups (**Figure 1b-c, i**). Direct grafting PFO molecules on the initial silica surface yields a large contact angle of about 98°, showing its hydrophobicity owing to the replacement of the hydroxyl groups with the hydrophobic PFO molecules. However, the moderate high contact angle suggests that low affinity hydrogen bonding interaction between the silica surfaces with water molecules still happens (**Figure 1b-c, ii**).

Alternately, the silica surface is first treated with oxygen plasma and then is grafted with the hydrophobic PFO molecules. For the silica with oxygen plasma treatment, the contact angle is 0°, suggesting the formation of a dense layer of surface hydroxyl groups (**Figure 1b-c, iii**). After grafting with the hydrophobic PFO molecules on the pretreated silica surface, the surface contact angle increases to about 122°. The high contact angle can be ascribed to the formation of a dense hydrophobic PFO molecular layer on the surface, which could inhibit both high affinity and low affinity water absorption to silica surface and thus prevent Q degradation of a silica microcavity (**Figure 1b-c, iv**).

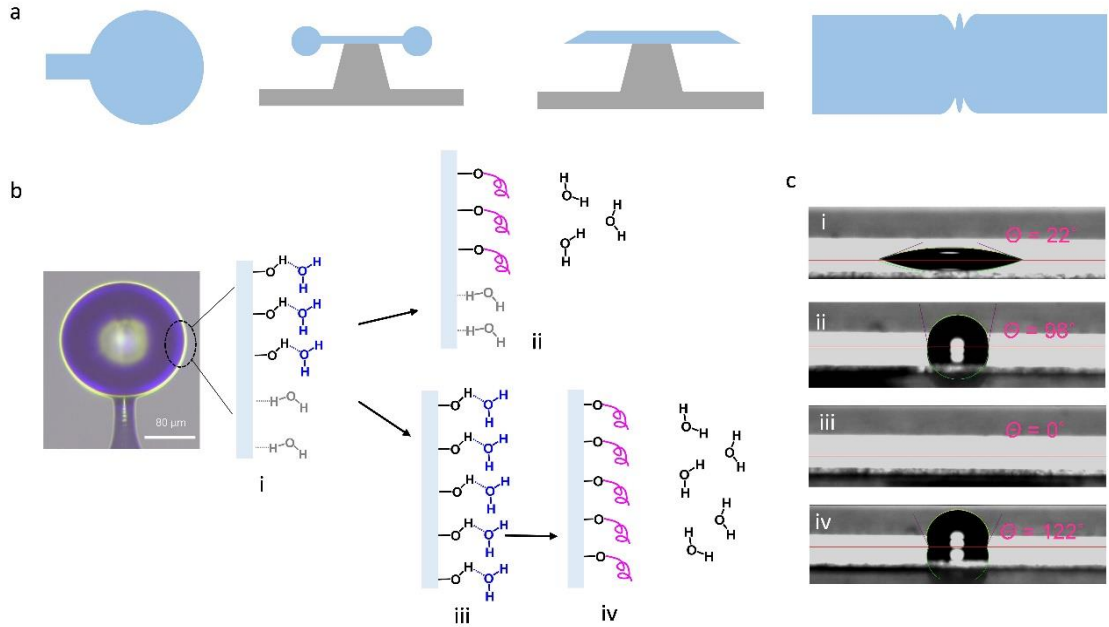

**Figure 1**. Surface chemistry of silica microcavities. (a) Illustration of typical silica microcavities (microsphere, microtoroid, microedge, microrod) with Q value excess 1x10$^8$. (b) Surface chemistry of a silica microsphere. An optical image of a silica microsphere with schematics showing the surface interaction with water molecules (i). The surface chemistry of silica grafted with PFO molecules without oxygen plasma pretreatment (ii), after oxygen plasma treatment (iii), and grafted with PFO molecules with oxygen plasma pretreatment (iv). The water molecules forming high affinity hydrogen bonds are showed in blue color, and the water molecules forming low affinity hydrogen bonds interaction are showed in grey color. The hydrophobic molecules are illustrated in pink color. (c) Optical images of water droplets on different silica surfaces obtained from an optical tensiometer. The measured surface contact angles are showed in each panel.

To evaluate the hydrophobic surface on the Q stabilities, bare and PFO-grafted silica (PFO-silica) microspheres with diameter of about 160 μm are tested in open air using a tapered optical fiber coupling method (**Figure 2a**). A continuous wave laser at 1550 nm is used for the optical test. Q values of the devices are obtained by fitting the resonance peak in transmission spectra in under-couple region (**Figure 2b**). The initial silica devices exhibit ultrahigh Q values of $1\times10^8$ to $5\times10^8$. The devices are then grafted with a monolayer of hydrophobic PFO molecules.

The Q values of the freshly prepared devices are monitored before and after deposition of the PFO molecules to investigate the direct influence of the surface molecules on Q values. Impressively, after grafting with the PFO molecules surface, the Q values maintain about 96% to 100% of that of the initial freshly prepared device (**Figure 2c**). The minimal influence of the surface PFO molecules on high Q values also indicates the grafted PFO layer is highly smoothness that has minimal surface scatting loss, which is consistent with the AFM results.

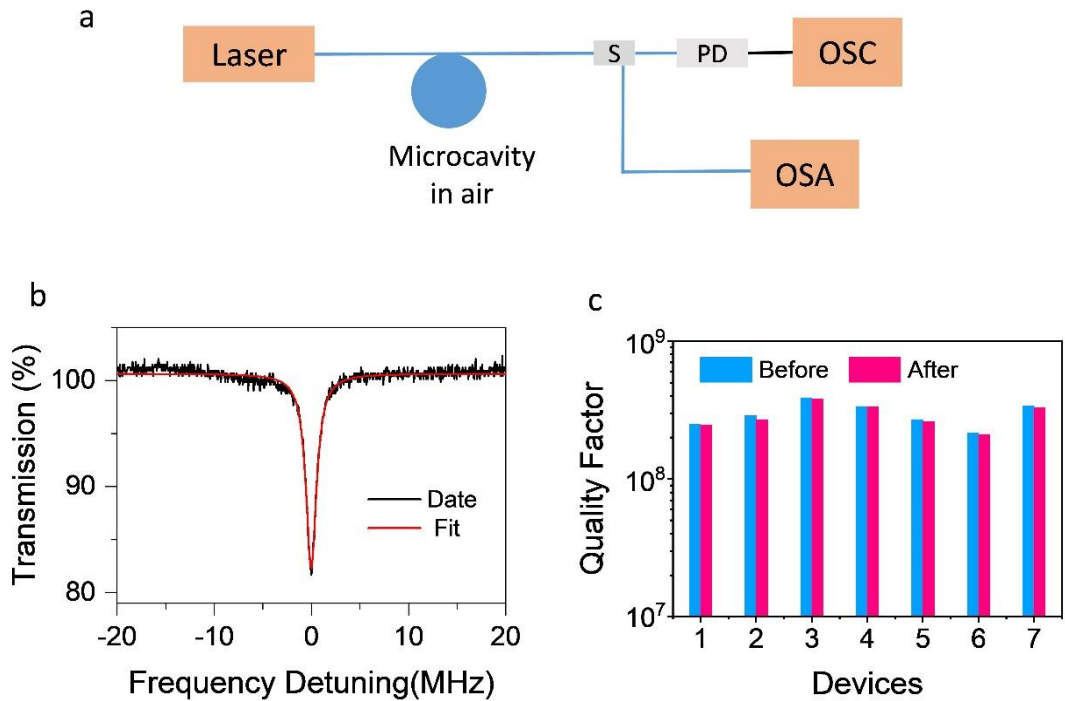

**Figure 2**. (a) Schematic of optical testing setup. Light (1550 nm) is coupled into a silica microcavity in air via an optical tapered fiber. The output is split by a fiber splitter (S), with 90% signal connected to an optical spectrum analyzer (OSA) and 10% signal connected to a photodetector (PD) followed by an oscilloscope (OSC). The test is conducted in air at room temperature. (b) A representative transmission spectrum of the device. (c) Comparison of Q values of the PFO silica devices before and after grafting with PFO molecules (seven devices are tested).

To evaluate the Q stabilities over time, all the silica microcavity devices are stored in open air with 35% or 75% humility at room temperature and monitored in the course of time. As expected, the Q values of initial silica devices severely decrease to about 10% of the original values after storage under 35% humility in air for 14 days (**Figure 3a**). However, the Q values of the PFO-silica devices show almost no degradation over times during storage under 35% (**Figure 3b**).or even 75% humility (**Figure 3c**), indicating that the surface hydrophobic PFO molecules can significantly improve the Q stabilities over time in air. The excellent Q stabilities can be ascribed to the formation of a dense molecule layer of PFO molecules via oxygen plasma pretreatment of silica, which is consolidated by direct comparing the Q stabilities of the PFO-silica devices with or without oxygen plasma pretreatment (**Figure 3d**). Both high affinity and low affinity hydrogen bonding interaction between water molecules and silica surface are prevented by the dense hydrophobic PFO monolayer.

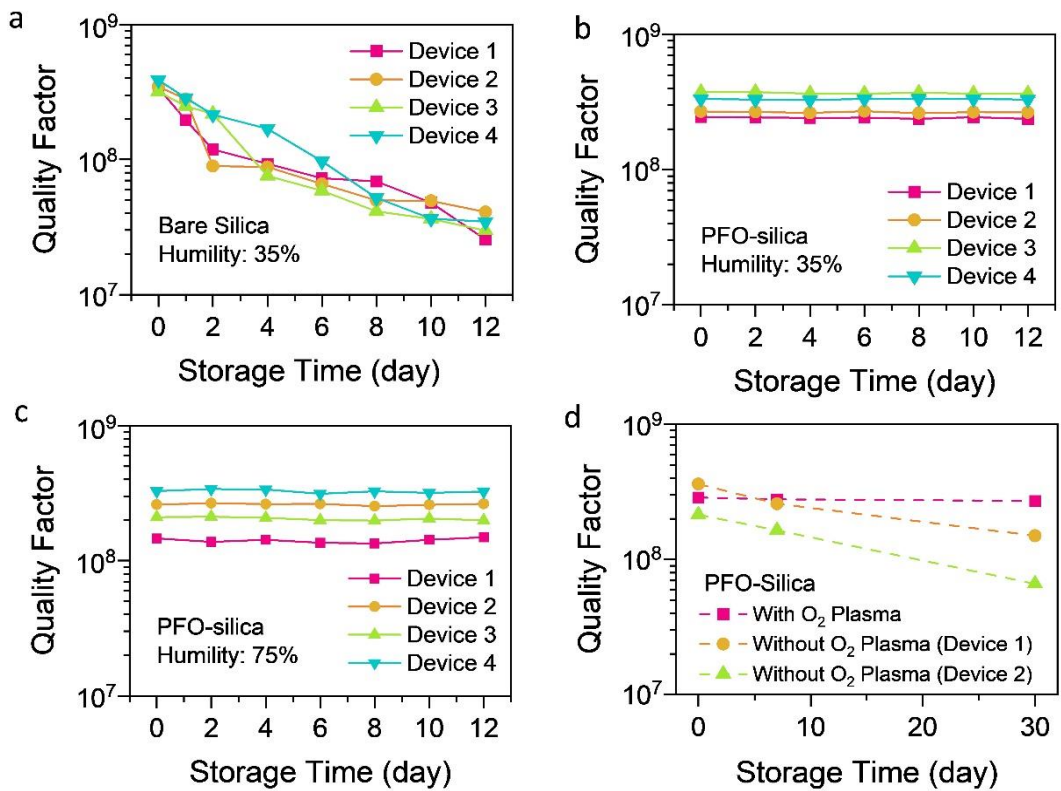

**Figure 3.** The Q stability over time of silica devices. (a) Bare silica and (b) the PFO-silica devices in open air with 35% humility. (c) PFO-silica devices in open air with 75% humility. (d) The PFO silica devices with or without oxygen plasma pretreatment in open air with 35% humility.

The influences of the dense surface hydrophobic PFO monomolecular layer on the devices nonlinear photonic performance are then evaluated. Both bare silica and PFO silica devices are stored in open air with 35% humidity. The devices are tested in the course of time in open air, pumped by a free running 1550 continuous laser via a tapered optical fiber waveguide. The output spectra are measured by an OSA, as shown in **Figure 2**.

**Figure 4a** shows the representative optical frequency comb spectra generated from the devices, with input power of about 1.4 mW. A comb spectra with line space of 11 free spectrum range (FSR) is observed. By fine tuning the pump wavelength, each comb line generates a subcomb and then merge together to form a full comb spectrum with one FSR. Combs spectra with different characteristics are observed under different pump modes (**Figure S5, supporting information**). For a PFO silica device and a freshly prepared bare silica device with similar size and Q factors, the generated frequency combs spectra show no obvious differences (**Figure 4b**). They both have a spectrum span of about 300 nm and a measured free spectrum range of about 3.0 nm near 1550 nm. It suggests that the hydrophobic PFO monomolecular layer has not affect the dispersion and optical mode of a device due to its small refractive index (1.39) and its negligible ultrathin thickness.

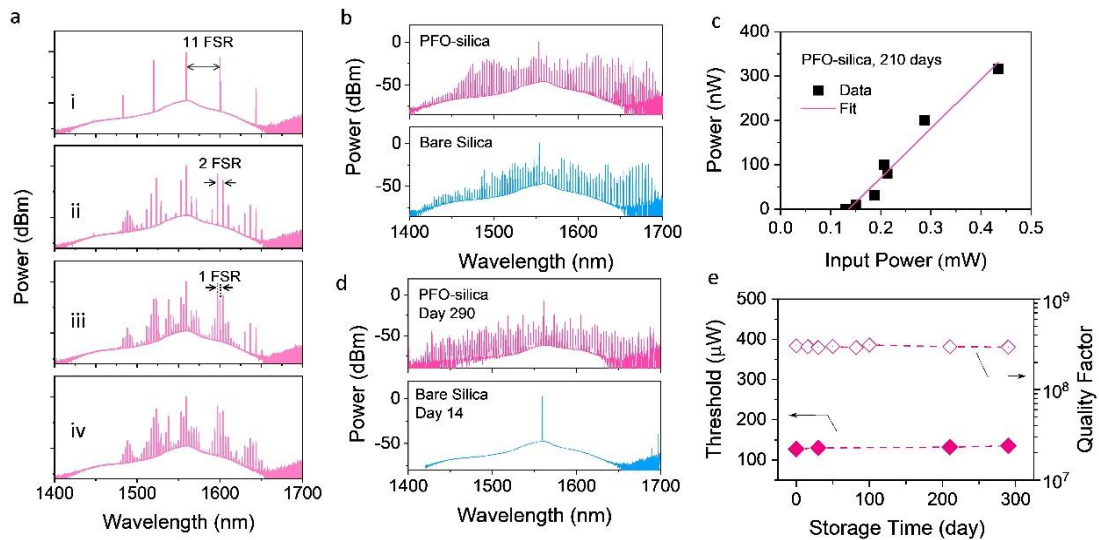

**Figure 4**. Frequency comb performance of the silica microcavities in open air with 35% humidity. (a) Evolution of a comb spectrum from a PFO-silica device via fine tuning of pump wavelength, with input power of about 1.4 mW. (b) Direct comparison of comb spectra of a PFO-silica and a bare silica under similar condition. (c) The dependency of comb line output power of a PFO-silica device on input power. (d) Direct comparison of comb spectra of a PFO-silica after storage for 290 days and a bare silica after storage for 14 days under similar condition. (e) The stabilities of comb threshold power and Q values of a PFO-silica device in the course of storage time.

Remarkably, PFO silica shows excellent stability for frequency comb generation. The threshold power is measured to be of about 117 µW (**Figure 4c**). Frequency comb spectra with similar threshold power and spectrum span are generated from the PFO silica device after storage in open air for 290 days (**Figure 4d**). Consistent with the Q values, the threshold values measured in the course of time show no obvious change (**Figure 4e**). However, for bare silica, no comb line is observed under the same input power after storage for 14 days. A measured threshold power high up to about 8.9 mW can be obtained by further increase input power (**Figure S6, supporting information**), which is governed by the theoretical quadratic dependency of threshold power to 1/Q. These results confirm that the dense surface hydrophobic PFO monomolecular layer endows the devices excellent stability in open air for sustainable frequency comb performance, which would greatly facilitate the long-term triggering and performance maintenance of silica-based soliton frequency combs.

**Conclusion**

In summary, we report a new type of ultrahigh Q silica microcavities that prevents the Q degradation over time, and enables their sustainable nonlinear photonic performances in air for long term. The silica device is grafted a dense hydrophobic monolayer of PFO molecules, which can effectively inhibit both high affinity and low affinity surface water absorption. Notably, the device shows exceptional long-term Q stability in air with 35% or even 75% humilities at room temperature, enabling the generation of optical frequency combs in air with sustainable output performances for long term. This work represents an effective and simple strategy to solve the Q degradation problem of silica microcavities to enable the long-term stable conversion of nonlinear frequencies. This approach is not limited to the specific device geometry used in the present work but is broadly applicable to any other silica-based resonant cavity devices. It paves the way for the practical applications of ultrahigh Q silica devices in next-generation nonlinear photonics.

**Methods**

**Device fabrication.** The bare silica microcavities are fabricated from optical fiber using a previous method. Briefly, the claddings are removed, the fibers are cleaned and exposed to a carbon dioxide laser to form silica microsphere cavities at the fiber ends. The diameters of the silica microsphere are approximately 160 μm. Then, the silica devices are treated by oxygen plasma, followed by deposition of *1H,1H,2H,2H*-perfluorooctyl trichlorosilane to the surface of the silica under vacuum for 8 min at room temperature, yielding a hydrophobic silica microcavity. For comparison, the bare silica devices without pretreated by oxygen plasma are directly functionalized with PFO trichlorisilanes under the same chemical vapor deposition condition. The fabricated devices are stored in air with 35% or 75% humility at room temperature.

**Surface characterization.** PFO monolayer is grafted on a thermal oxide silica surface of a silicon wafer for surface characterization. The surface chemical components of the silica devices are characterized by X-ray photoelectron spectroscopy (ESCALAB 250XI, ThermoFisher). The surface morphology and thickness of the PFO monolayer are characterized by using an atomic force microscope (Fastscan/Icon-AFM, Bruker). The contact angles of the surface are measured by using an optical tensiometer (Kino). Water droplet are placed on the silica surfaces and images of the drops are recorded. The static contact angles are then defined by fitting Young-Laplace equation around the droplet.

**Optical testing.** The devices for optical testing are tested in air at room temperature. The quality factors of the devices are characterized using a 1550 nm tunable laser (TOPTICA CTL 1550). Light is coupled into the devices using tapered optical fibers;

the power and coupling are precisely control for the calculation of the intrinsic Q. The transmission spectrum is recorded by an oscilloscope (Tektronix MDO3000). Multiple devices are fabricated and tested. The intrinsic Q is determined by acquiring the spectra over a range of coupling conditions.

For optical frequency comb measurement, the output signal from the tapered optical fiber is sent to a 10:90 splitter, with 90% signal connected to the photodetector and the optical spectrum analyzer (YOKOGAVA AQ6374).

**Acknowledgements**

This work are supported by the National Natural Science Foundation of China (Grant No. 62275152), the Shanghai Pujiang Program (Grant No. 20PJ1411600) and the Science and Technology Commission of Shanghai Municipality (Grant No. 20ZR1436400). We thank Xinyan Wang for the assistance of AFM measurement. We also acknowledge the support from ShanghaiTech University, the Soft Mater Nanofabrication Laboratory and the Analytical Instrumentation Center (Contract No. SPST-AIC10112914).

**Author contribution**

J. Xie and Y. Wang contributed equally to the work. Y. Wang conducted the devices fabrication and characterization. J. Xie conducted the optical measurement. H. Kang fabricated and characterized silica microrod cavities. J. Cheng fabricated and characterized silica microbubble cavities. X. Shen conceived the project. X. Shen wrote the manuscript with inputs from all the authors.

The authors declare no competing or conflicts of interest. Correspondence and requests for materials should be addressed to Xiaoqin Shen (shenxq@shanghaitech.edu.cn).

## Data Availability Statement

The data that support the plots with this paper and other findings of this study are available from the corresponding author upon reasonable request.

## Supporting Information

Additional Q stability characterization, XPS spectra, AFM results and additional comb performance of the devices.

## References

1. Vahala, K. J., Optical microcavities. *Nature* **2003,** *424* (6950), 839-846.
2. Kippenberg, T. J.; Holzwarth, R.; Diddams, S., Microresonator-based optical frequency combs. *Science* **2011,** *332* (6029), 555-559.
3. Diddams, S. A.; Vahala, K.; Udem, T., Optical frequency combs: Coherently uniting the electromagnetic spectrum. *Science* **2020,** *369* (6501), eaay3676.
4. Yao, L.; Liu, P.; Chen, H.-J.; Gong, Q.; Yang, Q.-F.; Xiao, Y.-F., Soliton microwave oscillators using oversized billion Q optical microresonators. *Optica* **2022,** *9* (5), 561-564.
5. He, L.; Özdemir, Ş. K.; Yang, L., Whispering gallery microcavity lasers. *Laser & Photonics Reviews* **2013,** *7* (1), 60-82.
6. Bai, Y.; Zhang, M.; Shi, Q.; Ding, S.; Qin, Y.; Xie, Z.; Jiang, X.; Xiao, M., Brillouin-Kerr soliton frequency combs in an optical microresonator. *Physical Review Letters* **2021,** *126* (6), 063901.
7. Liu, J.; Bo, F.; Chang, L.; Dong, C.-H.; Ou, X.; Regan, B.; Shen, X.; Song, Q.; Yao, B.; Zhang, W., Emerging material platforms for integrated microcavity photonics. *Science China Physics, Mechanics & Astronomy* **2022,** *65* (10), 104201.
8. Zhang, X.; Cao, Q.-T.; Wang, Z.; Liu, Y.-x.; Qiu, C.-W.; Yang, L.; Gong, Q.; Xiao, Y.-F., Symmetry-breaking-induced nonlinear optics at a microcavity surface. *Nature Photonics* **2019,** *13* (1), 21-24.
9. Lee, H.; Chen, T.; Li, J.; Yang, K. Y.; Jeon, S.; Painter, O.; Vahala, K. J., Chemically etched ultrahigh-Q wedge-resonator on a silicon chip. *Nature Photonics* **2012,** *6* (6), 369-373.
10. Shen, X.; Choi, H.; Chen, D.; Zhao, W.; Armani, A. M., Raman laser from an optical resonator with a grafted single-molecule monolayer. *Nature Photonics* **2020,** *14* (2), 95-101.
11. Liu, J.; Huang, G.; Wang, R. N.; He, J.; Raja, A. S.; Liu, T.; Engelsen, N. J.; Kippenberg, T. J., High-yield, wafer-scale fabrication of ultralow-loss, dispersion-engineered silicon nitride photonic circuits. *Nature communications* **2021,** *12* (1), 2236.

12. Boes, A.; Chang, L.; Langrock, C.; Yu, M.; Zhang, M.; Lin, Q.; Lončar, M.; Fejer, M.; Bowers, J.; Mitchell, A., Lithium niobate photonics: Unlocking the electromagnetic spectrum. *Science* **2023,** *379* (6627), eabj4396.

13. Lin, J.; Yao, N.; Hao, Z.; Zhang, J.; Mao, W.; Wang, M.; Chu, W.; Wu, R.; Fang, Z.; Qiao, L., Broadband quasi-phase-matched harmonic generation in an on-chip monocrystalline lithium niobate microdisk resonator. *Physical review letters* **2019,** *122* (17), 173903.

14. Wang, J.-Q.; Yang, Y.-H.; Li, M.; Hu, X.-X.; Surya, J. B.; Xu, X.-B.; Dong, C.-H.; Guo, G.-C.; Tang, H. X.; Zou, C.-L., Efficient frequency conversion in a degenerate χ (2) microresonator. *Physical Review Letters* **2021,** *126* (13), 133601.

15. Chang, L.; Xie, W.; Shu, H.; Yang, Q.-F.; Shen, B.; Boes, A.; Peters, J. D.; Jin, W.; Xiang, C.; Liu, S., Ultra-efficient frequency comb generation in AlGaAs-on-insulator microresonators. *Nature communications* **2020,** *11* (1), 1331.

16. Wang, C.; Li, J.; Yi, A.; Fang, Z.; Zhou, L.; Wang, Z.; Niu, R.; Chen, Y.; Zhang, J.; Cheng, Y., Soliton formation and spectral translation into visible on CMOS-compatible 4H-silicon-carbide-on-insulator platform. *Light: Science & Applications* **2022,** *11* (1), 341.

17. Del'Haye, P.; Diddams, S. A.; Papp, S. B., Laser-machined ultra-high-Q microrod resonators for nonlinear optics. *Applied Physics Letters* **2013,** *102* (22), 221119.

18. Wu, L.; Wang, H.; Yang, Q.; Ji, Q.-x.; Shen, B.; Bao, C.; Gao, M.; Vahala, K., Greater than one billion Q factor for on-chip microresonators. *Optics Letters* **2020,** *45* (18), 5129-5131.

19. Shen, X.; Beltran, R. C.; Diep, V. M.; Soltani, S.; Armani, A. M., Low-threshold parametric oscillation in organically modified microcavities. *Science advances* **2018,** *4* (1), eaao4507.

20. Jiang, B.; Zhu, S.; Ren, L.; Shi, L.; Zhang, X., Simultaneous ultraviolet, visible, and near-infrared continuous-wave lasing in a rare-earth-doped microcavity. *Advanced Photonics* **2022,** *4* (4), 046003-046003.

21. Zhang, H.; Tan, T.; Chen, H.-J.; Yu, Y.; Wang, W.; Chang, B.; Liang, Y.; Guo, Y.; Zhou, H.; Xia, H., Soliton Microcombs Multiplexing Using Intracavity-Stimulated Brillouin Lasers. *Physical Review Letters* **2023,** *130* (15), 153802.

22. Chen, J.-h.; Shen, X.; Tang, S.-J.; Cao, Q.-T.; Gong, Q.; Xiao, Y.-F., Microcavity nonlinear optics with an organically functionalized surface. *Physical review letters* **2019,** *123* (17), 173902.

23. Chen, D.; Kovach, A.; Shen, X.; Poust, S.; Armani, A. M., On-Chip Ultra-High-Q Silicon Oxynitride Optical Resonators. *ACS Photonics* **2017,** *4* (9), 2376-2381.

24. Jeong, D.; Kim, D.-G.; Do, I. H.; Lee, H., Hydrophobic passivation of ultra-high-Q silica wedge resonators using hexamethyldisilazane. *Optics Letters* **2021,** *46* (9), 2019-2022.

25. Hobza, P.; Sauer, J.; Morgeneyer, C.; Hurych, J.; Zahradnik, R., Bonding ability of surface sites on silica and their effect on hydrogen bonds. A quantum-chemical and statistical thermodynamic treatment. *The Journal of Physical Chemistry* **1981,** *85* (26), 4061-4067.

26. Henderson, M. A., The interaction of water with solid surfaces: fundamental aspects revisited. *Surface Science Reports* **2002,** *46* (1-8), 1-308.

27. Ulman, A., Formation and structure of self-assembled monolayers. *Chemical reviews* **1996,** *96* (4), 1533-1554.

Supporting information for

# Hydrophobic Silica Microcavities with Sustainable Nonlinear Photonic Performance

Jiadu Xie, Yang Wang, Hui Kang, Jinsong Cheng, Xiaoqin Shen*

School of Physical Science and Technology, ShanghaiTech University, Shanghai, 201210, China

Email: shenxq@shanghaitech.edu.cn

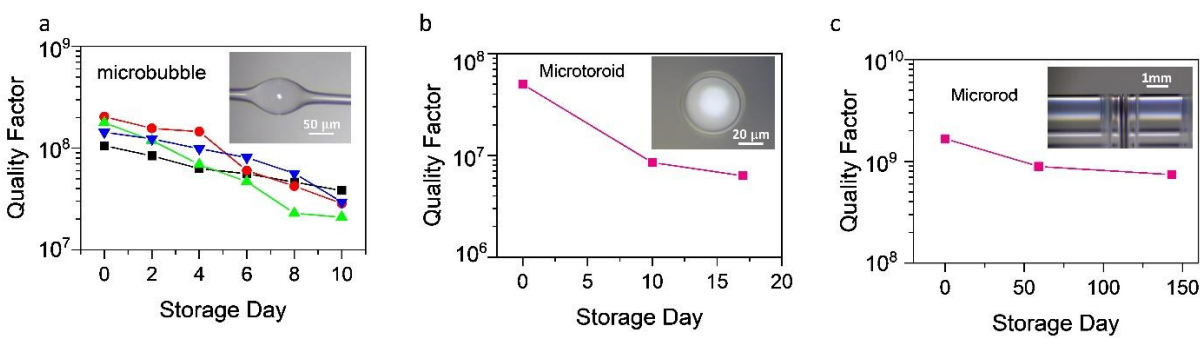

**Figure S1**. Q stabilities of different bare silica microcavities at ~ 1550 nm. (a) silica microbubbles, (b) silica microsphere, (c) silica microtoroid and (d) silica microtrod. All the bare silica devices experience Q degradation in air over times. Insets are the optical image of the fabricated devices.

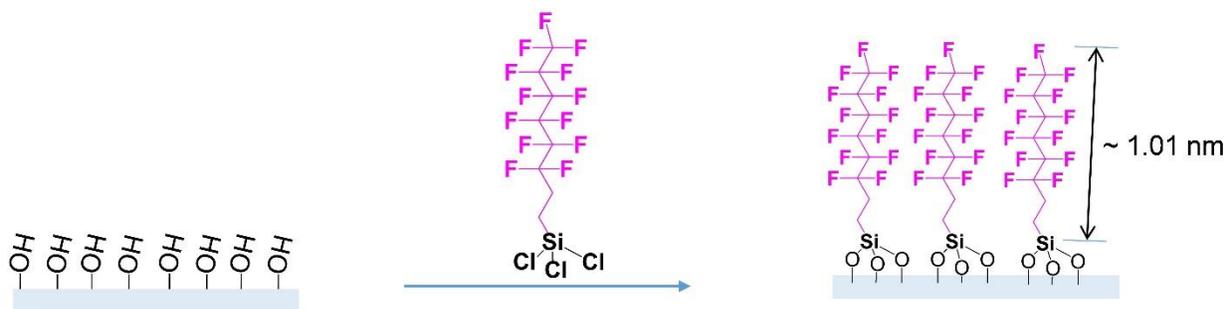

**Figure S2**. The schematics of the surface chemistry of the self-assembled monolayer of hydrophobic PFO molecules.

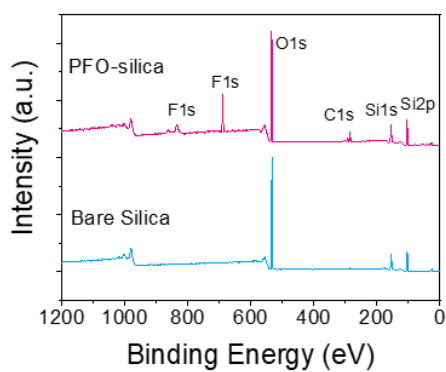

**Figure S3**. X-ray photoelectron spectroscopy (XPS) spectra of PFO-silica and bare silica surface. The binding energy peak at about 285 eV is the primary peak of C1s from the grafted PFO molecules. The binding energy peaks at about 687 and 832eV are the primary peak and its auger of F1s from the grafted PFO molecules and its auger, respectively. It confirms the formation of FPO molecular layer on silica surface.

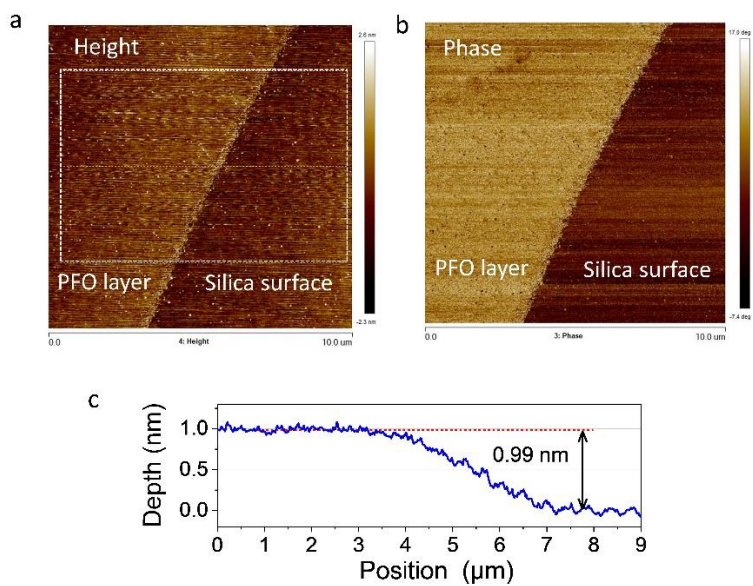

**Figure S4**. AFM measurement of PFO monolayer on silica surface. (a) and (b) are the height and phase mapping of a surface sample. The dashed area in (a) are chose to measure the step height of the PFO monolayer as shown in (c). The measured thickness of the PFO monolayer is of about 0.99 nm, which is consistent with the theoretical length of one PFO molecule (1.01 nm).

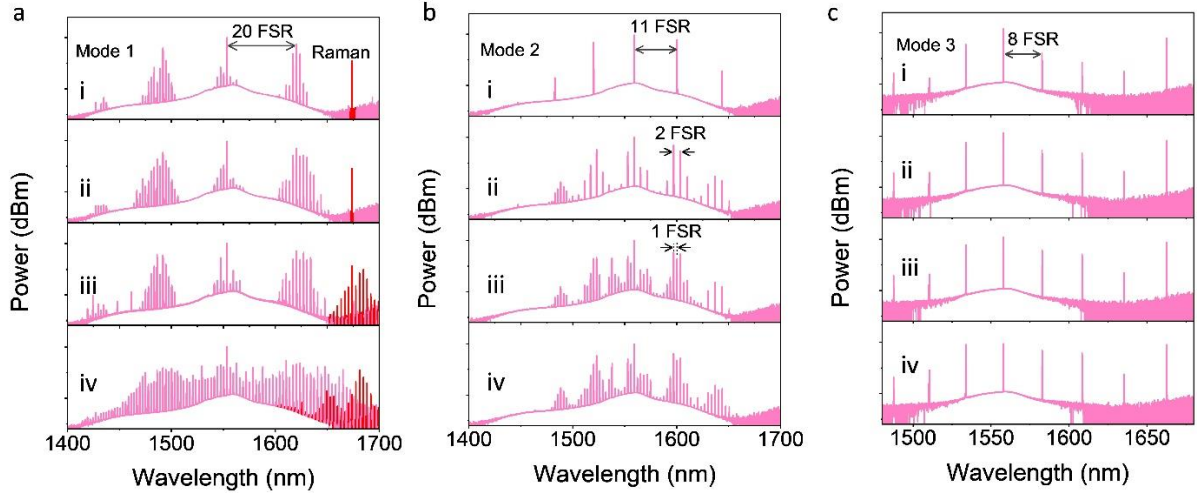

**Figure S5**. Optical frequency combs generation from a PFO-silica microcavity. (a) Under pump mode 1, a comb spectrum with a major line spacing of 20 FSR is generated. Subcombs with minor line spacing of 1 FSR and Raman lasing line is also observed. By red tuning the pump wavelength (from i to iii), the subcombs spectra expended and then merge together (iv), forming a comb spectrum (1 FSR line spacing) with spectra span of about 300 nm. A Raman comb (1 FSR lien spacing) is also observed. (b) Under pump mode 2, a comb spectrum with a major line of spacing of 11 FSR is generated (i). By red tuning the pump wavelength, subcombs with 2 FSR line spacing（ii）and then 1 FSR line spacing (iii ) are observed. The subcombs then merge together, forming a comb spectrum (1 FSR line spacing) with spectra span of about 180 nm (iv). (c) Under pump mode 3, a comb spectra with line spacing of 8 FSR is generated. By red tuning the pump wavelength (i to iv), the comb line power increase while no subcomb is observed.

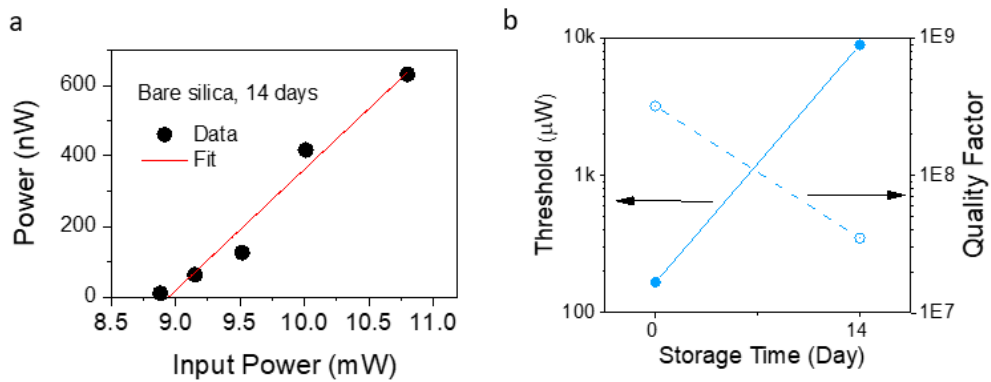

**Figure S6**. (a) The dependency of comb line output power on input power of a bare silica device after storage for 14 days. The threshold power of frequency comb generation is 8.9 mW. (b) The Q values and threshold powers of frequency comb generation of a bare silica device over times. After storage for 14 days, the Q value of the bare silica decrease from $3.2 \times 10^8$ to $3.5 \times 10^7$, leading to the increase of the threshold power from 167 μW to 8.9 mW.